\documentclass[aps,twocolumn,preprintnumbers,eqsecnum,prb]{revtex4}
\usepackage{amsfonts}
\usepackage[dvips]{graphicx}
\usepackage[T1]{fontenc}
\usepackage{amsmath,amsbsy,amssymb,graphics}

\begin{document}

\preprint{PHYSICAL REVIEW B \textbf{73}, 045432 (2006); 
Virtual Journal of Nanoscale Science \& Technology - Feb.6, 2006 Vol.13, Issue 5}
\title{Peculiar Width Dependence of the Electronic Property of Carbon
Nanoribbons}
\author{Motohiko Ezawa}
\affiliation{{}Department of Physics, University of Tokyo, Hongo 7-3-1, 113-0033, Japan }

\begin{abstract}
Nanoribbons (nanographite ribbons) are carbon systems analogous to carbon
nanotubes. We characterize a wide class of nanoribbons by a set of two
integers $\langle p,q\rangle $, and then define the width $w$ in terms of $p$
and $q$. Electronic properties are explored for this class of nanoribbons.
Zigzag (armchair) nanoribbons have similar electronic properties to armchair
(zigzag) nanotubes. The band gap structure of nanoribbons exhibits a valley
structure with stream-like sequences of metallic or almost metallic
nanoribbons. These sequences correspond to equi-width curves indexed by $w$.
We reveal a peculiar dependence of the electronic property of nanoribbons on
the width $w$.
\end{abstract}

\maketitle

\section{Introduction}

Nanometric carbon materials exhibit various remarkable properties depending
on their geometry\cite%
{Dresselhaus,Kroto,Iijima,Hamada,Appl,Mintmire,Wildoer,Jorio,Shima,Saito,Ouyang}%
. In particular, intensive research has been made on carbon nanotubes\cite%
{Iijima} in the last decade. Carbon nanotubes are obtained by wrapping a
graphene sheet into a cylinder form. The large interest centers their
peculiar electronic properties inherent to quasi-one-dimensional systems.

A similarly fascinating carbon system is a stripe of a graphene sheet named
nanographite ribbons, graphene ribbon or nanographene\cite%
{Kivelson,Tanaka,Dunlap,Fujita,Ajiki,Wakaba,Ryu,Duplock}. We call them 
\textit{carbon nanoribbons} in comparison with carbon nanotubes. They can be
manufactured by deposition of nanotubes or diamonds\cite%
{Murakami,Zhang,Shibayama,Affoune}. Experimental studies have begun only
recently\cite{Cancado,Niimi,Kobayashi,Matsui}. Nanoribbons have a higher
variety than nanotubes because of the existence of edges. Wide nanoribbons
with zigzag edges have been argued to possess the flat band and show edge
ferromagnetism\cite{Fujita}. Though quite attractive materials, their too
rich variety has made it difficult to carry out a systematic analysis of
carbon nanoribbons.

In this paper we make a new proposal to characterize a wide class of
nanoribbons by a set of two integers $\langle p,q\rangle $ representing edge
shape and width. The width $w$ is defined in terms of $p$ and $q$. We
present a systematic analysis of their electronic property in parallel to
that of nanotubes. Carbon nanotubes are regarded as a
periodic-boudary-condition problem while carbon nanoribbons are as a
fixed-boudary-condition problem. By calculating band gaps they are shown to
exhibit a variety of properties in electronic conduction, from metals to
typical semiconductors. Several sequences of metallic or almost metallic
(MAM) points are found in the valley of semiconducting nanoribbons. We
reveal a peculiar dependence of the electronic property of nanoribbons on
the width $w$. For instance, these sequences and equi-width curves become
almost identical for wide nanoribbons. We also point out that the
distribution of van-Hove singularities as a function of $w$ shows a peculiar
stripe pattern.

This paper is composed as follows. In section \ref{SecCCN} we characterize a
wide class of nanoribbons by a set of two integers $\langle p,q\rangle $ and
introduce the width $w$. In section \ref{SecElec}, making a numerical study,
we present an overview of the band gap structure for this class of
nanoribbons. In section \ref{SecComp} we compare nanoribbons with nanotubes.
Zigzag nanoribbons, being indexed by $\langle p,0\rangle $ with even $p$,
correspond to armchair nanotubes; armchair nanoribbons, indexed by $\langle
p,1\rangle $ with odd $p$, correspond to zigzag nanotubes. In section \ref%
{SecMet} we discuss sequences of metallic points developed in the valley of
semiconducting nanoribbons, where the metallic points on the principal
sequence are derived analytically. In section \ref{SecWid} we analyze
sequences of MAM points more in detail. The $n$-th sequence starts from the
metallic armchair nanotube $\langle 3n-1,1\rangle $. It approaches the
equi-width curve with $w=n$ for wide nanoribbons. In section \ref{SecEd} we
discuss edge effects. We take into account them in three ways: the
nonuniform site energy, the nonuniform transfer energy and the band filling
factor.

\begin{figure}[h]
\includegraphics[width=0.44\textwidth]{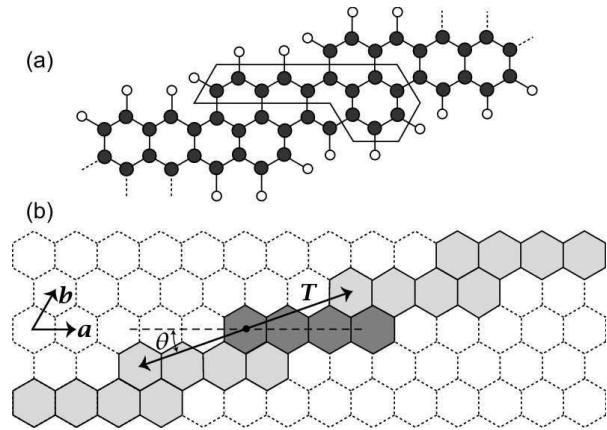}
\caption{{}(a) A typical structure of nanoribbons. A solid circle stands for
a carbon atom with one $\protect\pi $ electron, while an open circle for a
different atom such as a hydrogen.\ A closed area represents a unit cell. It
is possible to regard the lattice made of solid circles as a part of a
honeycomb lattice.\ (b) A nanoribon is constructed from a chain of $m$
connected carbon hexagons, as depicted in dark gray, and by translating this
chain by the translational vector $\mathbf{T}=\pm q\mathbf{a}+\mathbf{b}$
many times, as depicted in light gray, where $q<m$. A nanoribbon is indexed
by a set of two integers $\langle p,q\rangle $\ with $p=m-q$. Here we have
taken $m=4$, $q=2$, $p=2$. }
\label{FigDeco}
\end{figure}

\section{Classification of Carbon Nanoribbons}

\label{SecCCN}

A carbon nanoribbon is a one-dimensional aromatic compound. We have
illustrated a typical structure in Fig.\ref{FigDeco}(a). It consists of
carbon atoms of a honeycomb structure. A carbon on the edge is terminated by
a different atom such as a hydrogen so that it has no dangling bond. All
carbon atoms are connected by $\sigma $ bonds between sp$^{2}$ hybridized
orbitals of 2s, 2p$_{x}$ and 2p$_{y}$, providing with a framework of
honeycomb lattice. On the other hand, the $\pi $ bond is formed between two
2p$_{z}$ orbitals. The $\pi $ bands cross the Fermi energy, while the $%
\sigma $ bands are far away from it. Hence it is a good approximation to
take into account only $\pi $ electrons to investigate electronic properties
of nanoribbons. Each carbon atom has the complete shell and there is one
electron per atom.

Embedding them into a honeycomb lattice [Fig.\ref{FigDeco}(b)], we classify
nanoribbons as follows. An arbitrary lattice point on a honeycomb lattice is
described by the lattice vector 
\begin{equation}
\mathbf{R}=x\mathbf{a}+y\mathbf{b},
\end{equation}%
where $\mathbf{a}$ and $\mathbf{b}$ are primitive lattice vectors while $x$
and $y$ are integers [Fig.\ref{FigDeco}(b)]. First we take a basic chain of $%
m$ connected carbon hexagons, as depicted in dark gray. Second we translate
this chain by the translational vector 
\begin{equation}
\mathbf{T}=\pm q\mathbf{a}+\mathbf{b},
\end{equation}%
as depicted in light gray, where $q$ is an arbitrary integer $\left(
q<m\right) $. Repeating this translation many times we construct a
nanoribbon indexed by a set of two integers $\langle p,q\rangle $, where $%
p=m-q$. In what follows we analyze the class of nanoribbons generated in
this way.

\begin{figure}[h]
\includegraphics[width=0.4\textwidth]{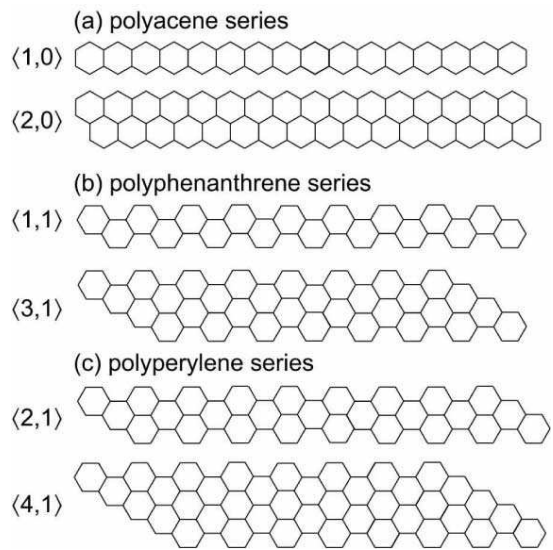}
\caption{{}The geometric configuration of (a) the polyacene series having
zigzag edges, (b) the polyphenanthrene series having armchair edges, and (c)
the polyperylene series having armchair edges.}
\label{Exa}
\end{figure}

The indices $p$ and $q$ specify the type of nanoribbons. The case with $q=0$
represents a zigzag nanoribbon with zigzag edge, while the case with $q=1$
represents an armchair nanoribbon with armchair edge. The nanoribbons with $%
\langle 1,0\rangle $, $\langle 1,1\rangle $ and $\langle 2,1\rangle $ are
known as polyacene, polyphenanthrene and polyperynaphthalene\cite%
{Kivelson,Tanaka,Murakami,Yudasaka,Affoune}, respectively [Fig.\ref{Exa}]. 

We propose to define the width of the nanoribbon by%
\begin{equation}
w=2m\sin \theta =\frac{2\left( p+q\right) }{\sqrt{3\left( 2q+1\right) ^{2}+9}%
},  \label{WidthParam}
\end{equation}%
where $\theta $ is the angle between the basic chain and the translational
vector, 
\begin{equation}
\tan \theta =\frac{\sqrt{3}}{2q+1},
\end{equation}%
as in Fig.\ref{FigDeco}(b). There is a freedom to normalized the width. We
have normalized the width of an armchair nanoribbon indexed by $\left(
p,1\right) $ to be 
\begin{equation}
w=\frac{p+1}{3}  \label{WidthArm}
\end{equation}%
by the reason that becomes clear in Section \ref{SecWid}. Then, the width of
zigzag nanoribbons indexed by $\left( p,0\right) $ becomes 
\begin{equation}
w=\frac{p}{\sqrt{3}}.  \label{WidthZig}
\end{equation}%
The width of a nanoribbon corresponds to the diameter of a nanotube. Though
a nanoribbon is specified by two integers, we expect that the electronic
property of wide nanoribbons is mainly controlled by the width $w$.

The above classification rule is similar to that of nanotubes based on the
chiral vector or the rolling up vector\cite{Saito,Ouyang,Seifert2}, but it
is clearly different since some nanoribbons can not be rolled up into
nanotubes. We discuss the correspondence between nanoribbons and nanotubes
in Section \ref{SecComp}.

\section{Electronic structure of nanoribbons}

\label{SecElec}

We calculate the band structure of nanoribbons based on the nearest-neighbor
tight-binding model, which has been successfully applied to the studies of
carbon nanotubes\cite{Saito}.

The tight-binding Hamiltonian is defined by%
\begin{equation}
H=\sum_{i}\varepsilon _{i}c_{i}^{\dagger }c_{i}+\sum_{\left\langle
i,j\right\rangle }t_{ij}c_{i}^{\dagger }c_{j},  \label{HamilTB}
\end{equation}%
where $\varepsilon _{i}$ is the site energy, $t_{ij}$ is the transfer
energy, and $c_{i}^{\dagger }$ is the creation operator of the $\pi $
electron at the site $i$. The summation is taken over the nearest neighbor
sites $\left\langle i,j\right\rangle $. In the case of nanotubes, constant
values are taken for $\varepsilon _{i}$ and $t_{ij}$ owing to their
homogeneous geometrical configuration. Furthermore, since there exists one
electron per one carbon, the band-filling factor is 1/2. In the case of
nanoribbons, on the contrary, they would be modified by the existence of the
edges. (a) The site energy would be modified by the difference of
electronegativity of X, where X represents a different atom such as
hydrogen. (b) The transfer energy would be modified by a possible lattice
distortion near the edge. (c) The band-filling factor would be modified due
to the dipole moment of C-X bonds. It is our basic assumption that the
carbon nanoribbon can be analyzed based on this Hamiltonian together with
these three modifications.

It is convenient to take a unit cell as shown in Fig.\ref{FigDeco}(a). There
are $\left( 4q+2p+2\right) $ carbon atoms in a unit cell of the nanoribbon
indexed by $\langle p,q\rangle $, as implies that the proper functions of
the Hamiltonian $H$ consist of $\left( 4q+2p+2\right) $ Bloch wave functions 
$\left\vert \psi _{i}\right\rangle $. We take overlap integrals as 
\begin{equation}
\left\langle \psi _{j}\right. \left\vert \psi _{i}\right\rangle =\delta
_{ij},
\end{equation}
where $\delta _{ij}$ is Kronecker's delta. The $\pi $ bands of nanoribbon
are derived from the Hamiltonian $H\left( k;p,q\right) $ with $k$ the
crystal momentum, which is a $\left( 4q+2p+2\right) \times \left(
4q+2p+2\right) $ matrix. The band structure is determined by 
\begin{equation}
\det \left[ \varepsilon I-H\left( k;p,q\right) \right] =0,
\label{EugenProbl}
\end{equation}%
where $I$ is a unit matrix due to the overlap integral.

\begin{figure}[t]
\includegraphics[width=0.47\textwidth]{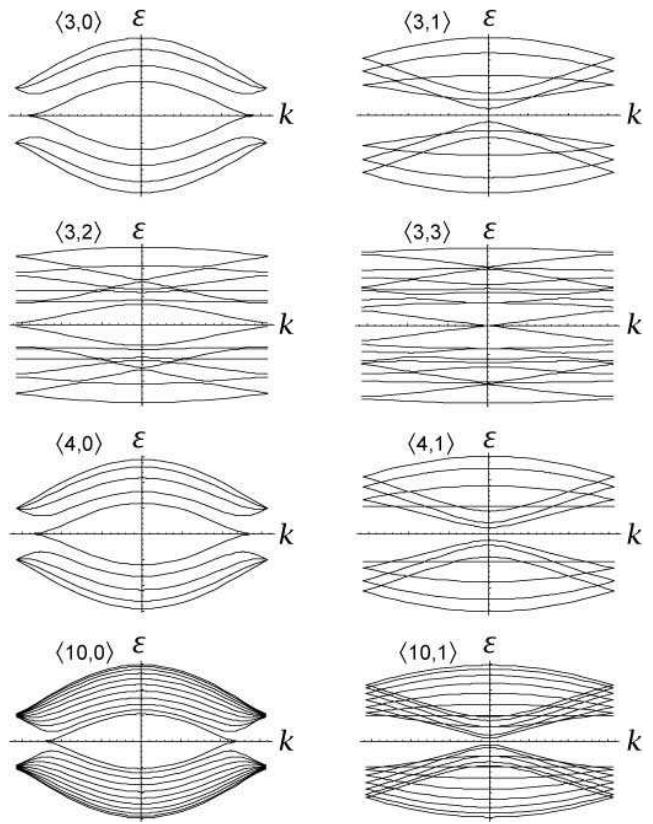}
\caption{The band structure of nanoribbons. The horizontal axis is the
crystal momentum $k$, $-\protect\pi <k\leq \protect\pi $, while the vertical
axis is the energy $\protect\varepsilon $, $-3\left\vert t\right\vert \leq 
\protect\varepsilon \leq 3\left\vert t\right\vert $ with $|t|=3.033$eV. The
band structure depends strongly on the index $q$ for fixed $p$, but depends
on the index $p$ only weakly for fixed $q$.}
\label{FigBandGap}
\end{figure}

Though the nanoribbon may have different values of $\varepsilon _{i}$ and $%
t_{ij}$ for atoms on the edge from the others, as we have remarked, the
difference is expected to be quite small. We first neglect the difference.
Namely, we take the transfer energy to be $t$ between all the nearest
neighbor sites and otherwise to be $0$. It is generally taken\cite{Saito} as 
$t=-3.033$eV. We also neglect the site energy term in the Hamiltonian (\ref%
{HamilTB}) by taking all the site energy $\varepsilon _{i}$ equal. We
discuss how the gap structure is modified by edge corrections in Section \ref%
{SecEd}.

\begin{figure}[t]
\includegraphics[width=0.45\textwidth]{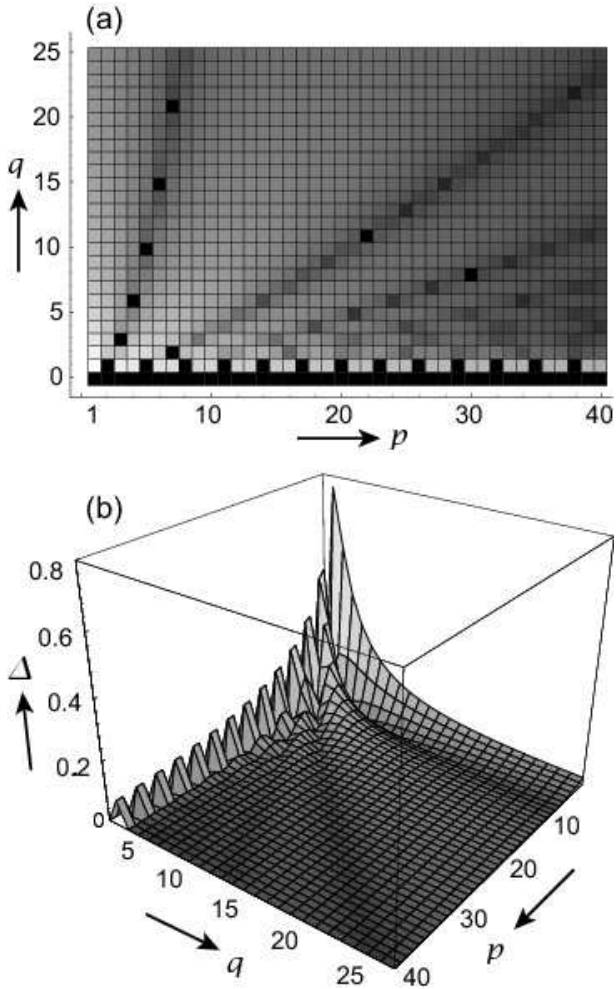}
\caption{{}The band gap structure of nanoribbons. (a) The horizontal and
vertical axes represent the indices $p$ and $q$, respectively. Magnitudes of
band gaps are represented by gray squares. Dark (light) gray squares
represent small (large) gap semiconductors. Especially, metallic states are
represented by black squares. (b) A bird's eye view. The vertical axis is
the energy gap $\Delta $ in unit of $|t|=3.033$eV. Nanoribbons make a valley
structure with stream-like sequences of metallic points in the $pq$ plane.}
\label{Valley1}
\end{figure}

We have solved the eigenvalue problem (\ref{EugenProbl}) numerically for $%
\varepsilon $. Typical band structures are shown in Fig. \ref{FigBandGap}.
As is seen in the figures, band structures depend strongly on the parity of $%
q$, but only weakly on $p$. We also find the following: (a) For metallic
nanoribbons, the Fermi point of even $q$ is at $k=\pi $, and that of odd $q$
is at $k=0$. (b) For semiconducting nanoribbons, the band gap minimum of
even $q$ is at $k=\pi $, and that of odd $q$ is at $k=0$.

As a main result we display an overview of the band gap structure of
nanoribbons in Fig. \ref{Valley1}. Gapless states (represented by black
squares) are metallic, and gapfull states (represented by all other squares)
are semiconducting. There are a variety of semiconducting nanoribbons from
almost gapless ones (represented by dark gray squares) to large gapfull ones
(represented by light gray squares). We observe clearly three emergence
patterns of metallic points: (a) Metallic points $\langle p,0\rangle $ for
all $p$. (b) Metallic points $\langle p,1\rangle $ with $p=2,5,8,11,\cdots $%
. (c) Several sequences of metallic points on "streams" in valleys. We
discuss (a) and (b) in Section \ref{SecComp}, and (c) in Sections \ref%
{SecMet} and \ref{SecWid}.

\section{Nanoribbons versus Nanotubes}

\label{SecComp}

It is observed that nanoribbons indexed by $\langle p,0\rangle $ are
metallic for all $p$, which are in the polyacene series with zigzag edges
[Fig. \ref{Exa}(a)]. Nanoribbons indexed by $\langle p,1\rangle $ with $%
p=2,5,8,11,\cdots $ are found to be also metallic, which have armchair edges
[Fig. \ref{Exa}(b) and (c)]. This series has period $3$, as is a
reminiscence of the classification rules familiar for nanotubes. The
classification rule says that a nanotube is metallic \ when $n_{1}-n_{2}$ is
an integer multiple of $3$, and otherwise semiconducting, where $\left(
n_{1},n_{2}\right) $ is a chiral vector of the nanotube.

Let us explore the correspondences more in detail. There is a group of
nanoribbons each of which is constructed as a development of a nanotube by
cutting it along the translational vector. For example, a zigzag nanoribbon $%
\left( q=0\right) $ with even $p$ may be regarded as a development of the
armchair nanotube whose chiral vector is $\left( p/2,p/2\right) $. All
zigzag nanoribbons are metallic, as corresponds to the fact that all
armchair nanotubes are metallic. As another example, an armchair nanoribbon $%
\left( q=1\right) $ with odd $p$ may be regarded as a development of a
zigzag nanotube whose chiral vector is $\left( \left( p+1\right) /2,0\right) 
$. Armchair nanoribbons are metallic with period of $3$, as corresponds to
the fact that metallic zigzag nanotubes emerge by period of $3$.

The correspondence between these metallic points may be explained by the
absence of spiral currents in carbon nanotubes. Namely, currents flowing
along the axis of nanotubes are not affected by cutting along the axis.
There is no direct correspondence between chiral nanotubes and nanoribbons
for $q\geq 2$.

\section{Sequence of metallic Nanoribbons}

\label{SecMet}

It is remarkable that there are new series of discrete metallic points on
one-dimensional curves in Fig.\ref{Valley1}. These curves look like streams
in valleys. The prominent ones are at $\langle 1,0\rangle $, $\langle
2,1\rangle $, $\langle 3,3\rangle $, $\langle 4,6\rangle $, $\langle
5,10\rangle $, $\langle 6,15\rangle $, $\cdots $. We regard them to form the
principal sequence of metallic points of nanoribbons. There are also
sequences of metallic points on higher curves.

We are able to derive the principal sequence analytically. We start with the
observation that the density of states at the Fermi energy is $D(\varepsilon
_{\text{F}})\neq 0$, if the band structure of nanoribbons are gapless;
otherwise, $D(\varepsilon _{\text{F}})=0$. This follows from the reflection
symmetry around $\varepsilon =0$ and the existence of one electron per one
atom. The reflection symmetry is due to a bipartite lattice structure of
graphite\cite{Lieb}. Consequently, to investigate the metallic points, it is
enough to analyze%
\begin{equation}
\det \left[ H\left( k;p,q\right) \right] =0.  \label{detH}
\end{equation}%
We have $D(\varepsilon _{\text{F}})\neq 0$ if this equation has a solution,
and the nanoribbon is gapless; otherwise it is gapfull. We find simple
structures at $q=0$ and for $q\geq p-1$.

At $q=0$, the determinant (\ref{detH}) is explicitly calculated as%
\begin{equation}
\det \left[ H\left( k;p,0\right) \right] =\left( -1\right)
^{p+1}t^{2p+2}\left( 2\cos \frac{k}{2}\right) ^{2p+2}.
\end{equation}%
It vanishes for any $p$ at $k=\pi $. Because of this, every zigzag
nanoribbon is metallic. It follows that%
\begin{equation}
\lim_{p\rightarrow \infty }\det \left[ H\left( k;p,0\right) \right] \left\{ 
\begin{array}{cc}
=0 & \frac{2\pi }{3}<\left\vert k\right\vert \leq \pi \\ 
\neq 0 & \left\vert k\right\vert \leq \frac{2\pi }{3}%
\end{array}%
\right. .
\end{equation}%
We have thus verified analytically that the flat band emerges for $\frac{%
2\pi }{3}<\left\vert k\right\vert \leq \pi $ when the width of nanoribbons
is wide enough, as confirms a previous numerical result\cite{Fujita}, where
the flat band has been argued to lead to edge ferromagnetism.

For $q=1$ and $p=3n-1$ with integer $n$, the determinant (\ref{detH}) has a
factor such that%
\begin{equation}
\det \left[ H\left( k;p,0\right) \right] \propto \sin ^{2}\frac{k}{2},
\end{equation}%
and the nanoribbon is found to be gapless at $k=0$.

For $q\geq p-1$, the determinant (\ref{detH}) is calculated as%
\begin{align}
t^{-\left( 4q+2p+2\right) }& \left( -1\right) ^{p+1}\det \left[ H\left(
k;p,q\right) \right]  \notag \\
=& p\left( p+1\right) \cos 2k  \notag \\
& +\left( -1\right) ^{q}\left( p^{2}+p+2\right) \left( p+q+1\right) \cos k 
\notag \\
& +\left( p+q+1\right) ^{2}+\frac{p^{2}\left( p+1\right) ^{2}}{4}+1.
\end{align}%
We can prove that $\det \left[ H\left( k;p,q\right) \right] =0$ for a
certain $k$ provided that%
\begin{equation}
q=\frac{p\left( p-1\right) }{2}.  \label{MaginNumbe}
\end{equation}%
Nanoribbons are metallic on the points $\langle p,q\rangle $ with integers $%
p $ and $q$ with (\ref{MaginNumbe}). They constitute the principal sequence
of metallic points.

It is hard to solve $\det \left[ H\left( k;p,q\right) \right] =0$
analytically for $q<p-1$, though the existence of solutions is clear by
numerical analysis as in Fig.\ref{Valley1}(a). In this figure there are only
three metallic points; the two points are $\langle 7,2\rangle $ and $\langle
22,11\rangle $ on the second sequence, and the last point is $\langle
30,8\rangle $ on the third sequence.

Metallic points on higher sequences are quite curious in this respect. We
cannot tell how they arise systematically. However, it may be useless to
make efforts to distinguish between metallic and tiny-gap semiconductor too
seriously, since the simple tight-binding model we have used will not be
accurate enough to predict completely vanishing band gaps. Nevertheless the
valley structure with several "streams" will be a significant feature. Hence
it is more interesting how the sequences made of MAM (metallic or almost
metallic) points are located in valleys.

\section{Width-Dependence of Nanoribbons}

\label{SecWid}

We have defined the width $w$ of a nanoribbon by the formula (\ref%
{WidthParam}). We now argue that the sequences of MAM points are indexed by
this width. 
\begin{figure}[h]
\includegraphics[width=0.4\textwidth]{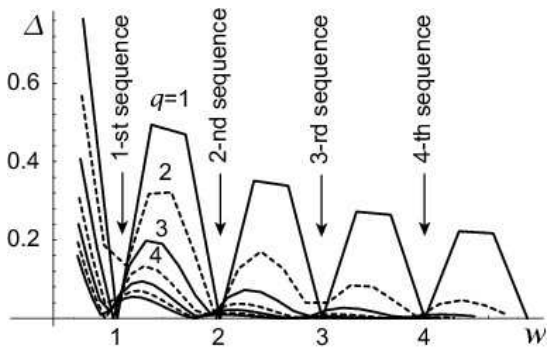}
\caption{{}The band gaps $\Delta $ in unit of $|t|=3.033$eV as a function of
the width $w$. The band gap behaves inversely to $w$. Band gaps take local
minima almost at the same values of the width $w$ for any $q$. }
\label{WB}
\end{figure}

First, in Fig.\ref{WB} we have depicted the band gap as a function of the
width $w$ for each fixed value of $q$. The band gap behaves inversely to $w$%
. This reminds us that the band gap of a carbon nanotube is inverse
proportion to the radius. The characteristic feature is that all band gaps
with different $q$ take local minima almost at the same values of width $w$, 
$w=1,2,3,\cdots $. It indicates that nanoribbons with similar width share
qualitatively the same electronic property. The first local minimum
corresponds to the principal sequence, which may be regarded as an extension
of the armchair nanoribbon indexed by $\langle 2,1\rangle $. In the same way
the $n$-th sequence may be regarded as an extension of the armchair
nanoribbon with $\langle 3n-1,1\rangle $.

\begin{figure}[h]
\includegraphics[width=0.4\textwidth]{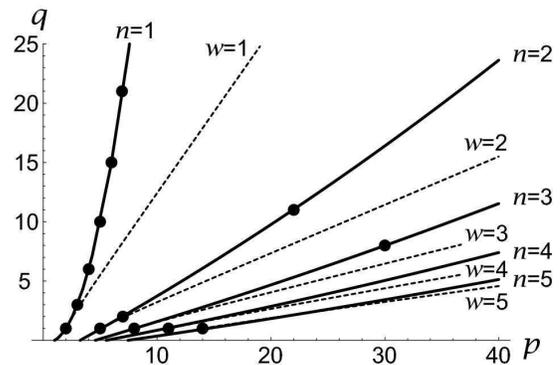}
\caption{{}Illustration of metallic points, sequences and equi-width curves.
Metallic points are denoted by solid circles. Solid curves represent
sequences of MAM points, while dotted curves represent the points $\langle
p,q\rangle $ possessing the same width. The $n$-th sequnce is tangent to the
equi-width curve with $w=n$ at $q=1$. These two curves become almost
identical for sufficiently wide nanoribbons.}
\label{Mosikiz}
\end{figure}

\begin{figure}[h]
\includegraphics[width=0.4\textwidth]{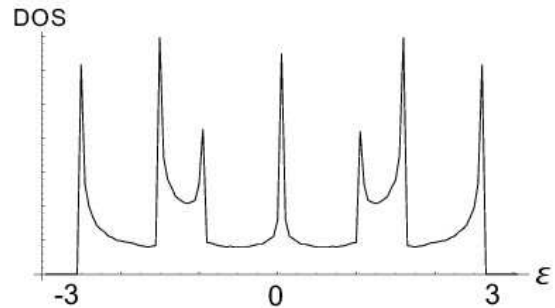}
\caption{The density of state (DOS) of the $\left\langle 1,0\right\rangle $
nanoribbon. The horizontal axis is the energy $\protect\varepsilon $. There
are many van-Hove singularity because of one-dimensional structure of
nanoribbon.}
\label{DOS}
\end{figure}

Solving eq.(\ref{WidthParam}) for $p$, we have%
\begin{equation}
p=-q+\frac{w}{2}\sqrt{3\left( 2q+1\right) ^{2}+9}.
\end{equation}%
In Fig.\ref{Mosikiz} we show the curves described by this equation and the
sequences of MAM points. It is observed that the $n$-th sequence is almost
tangent to the equi-width curve with $w=n$ at $q=1$. We are able to
associate the sequences with the equi-width curves in this way. These two
curves become almost identical for sufficiently wide nanoribbons. This
result may be understood as follows. In the case of a continuous ribbon with
no lattice structure, the only parameter is the width and the electronic
property is determined by this parameter. We present another indication that
the width $w$ is an interesting parameter. We calculate the density of state
of an arbitrary $\left\langle p,q\right\rangle $ nanoribbon numerically [Fig.%
\ref{DOS}]. There are many van-Hove singularities just as in nanotubes
because nanoribbons are also one-dimensional compounds. The global structure
of the density of state is determined by van-Hove singularities.\ These
peaks can be measured experimentally by Raman scattering\cite%
{Jorio,Rao,Cancado,Kim}.

\begin{figure}[h]
\includegraphics[width=0.47\textwidth]{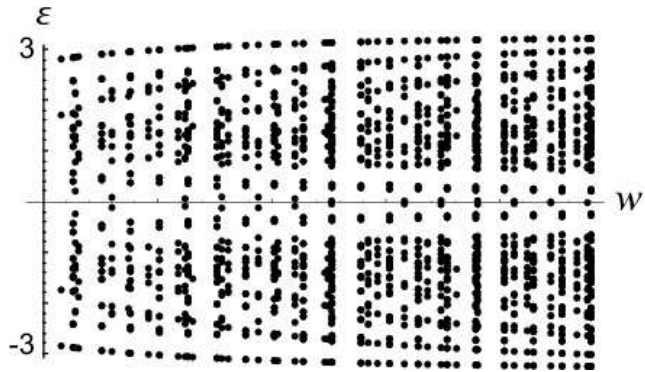}
\caption{{}Plot of van-Hove singularities in the $w$-$\protect\varepsilon $
plane. For a given $\left\langle p,q\right\rangle $ nanoribbon, we calculate
the width $w$ and the energies $\protect\varepsilon $ at which van-Hove
singularities develop. We have plotted the points $(w,\protect\varepsilon )$
for $q=1,2,3,4$ and for all $p$ in the region $w\leq 3$. A stripe pattern\
is manifest. }
\label{peak}
\end{figure}

We calculate the energies $\varepsilon $ at which van-Hove singularities
develop due to the local band flatness at $k=0$ for various $\left\langle
p,q\right\rangle $ nanoribbons. Note that the optical absorption is dominant
at $k=0$ because the dispersion relation $\varepsilon =ck$ with $c$ the
light velocity. On the other hand the width $w$ is determined by $p$ and $q$
as in (\ref{WidthParam}). We show the energy $\varepsilon $ of this peak as
a function of $w$ in Fig.\ref{peak}. A peculiar stripe pattern\ is manifest
there. In particular, the maximum and minimum values take almost the same
values $\pm 3\left\vert t\right\vert $, reflecting the electronic property
of a graphite\cite{Saito}. The fact that there are on smooth curves justify
a physical meaning of the width $w$. This stripe pattern would be accessible
experimentally by way of Raman scattering.

\section{Edge corrections}

\label{SecEd}

\begin{figure}[t]
\includegraphics[width=0.4\textwidth]{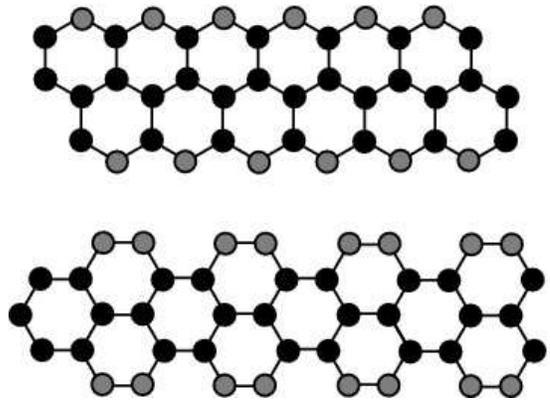}
\caption{Illustraion of edge carbons and bulk carbons. A gray circle denotes
an edge carbon connected by two carbons and a hydrogen. A black circle
denotes a bulk carbon connected by three carbons. The energy and the
transfer energy associated with edge carbons are set to be $\protect%
\varepsilon _{\text{edge}}$ and $t_{\text{edge}}$, respectively. Hydrogen
atoms are omitted in this figure.}
\label{FigDop}
\end{figure}

We finally study how the gap structure is modified by the existence of edges
in a carbon nanoribbon. All carbon atoms in a carbon nanotube are equivalent
in the sense that each of them is always surrounded by three carbon atoms.
In contrast, this is not the case for a nanoribbon, where a carbon on the
edge has less neighboring carbon atoms. The presence of C-X bonds introduces
carbon atoms on edge with a different nature. We assume that the edge
effects can be taken into account by modifying the band-filling factor, the
transfer energy $t_{ij}$ and the site-energy $\varepsilon _{i}$ in the
Hamiltonian (\ref{HamilTB}). We take $t_{ij}=t_{\text{edge}}$ and $%
\varepsilon _{i}=\varepsilon _{\text{edge}}$ for those associated with edge
carbons, and $t_{ij}=t_{\text{bulk}}$ and $\varepsilon _{i}=\varepsilon _{%
\text{bulk}}$ for bulk carbons [Fig.\ref{FigDop}]. 

First, the band-filling factor is affected by the dipole moment of C-X
bonds. The change of electron number is given at most by the number of C-X
bonds and is small for wide nanoribbons. This effect does not modify the
band structure, but only changes the occupancy of the band. As a result,
semiconducting nanoribbons tend to become metallic nanoribbons.

\begin{figure}[h]
\includegraphics[width=0.45\textwidth]{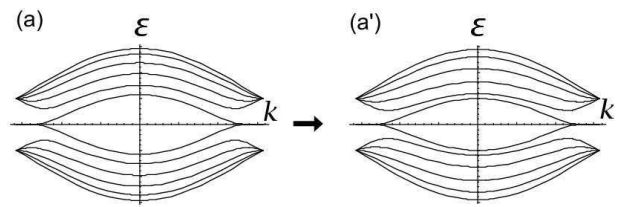}
\caption{{}(a) The original band structure of the $\langle 5,0\rangle $
nanoribbon. (a') The band structure of the nanoribbon with the choice of $t_{%
\text{edge}}=0.7t_{\text{bulk}}$. The difference is hardly recognizable.}
\label{EdgeBandT}
\end{figure}

Second, the transfer energy is affected by the change of the distance
between two carbons near the edge. The distortion does not break the
inversion symmetry of the band structure. Recalculating the band structure
of various nanoribbons, we have found that the difference is hardly
recognizable. We show the band structure of the $\left\langle
5,0\right\rangle $ nanoribbon in Fig.\ref{EdgeBandT}, assuming a large
correction such that $t_{\text{edge}}=0.7t_{\text{bulk}}$ to make the
difference recognizable.

\begin{figure}[h]
\includegraphics[width=0.45\textwidth]{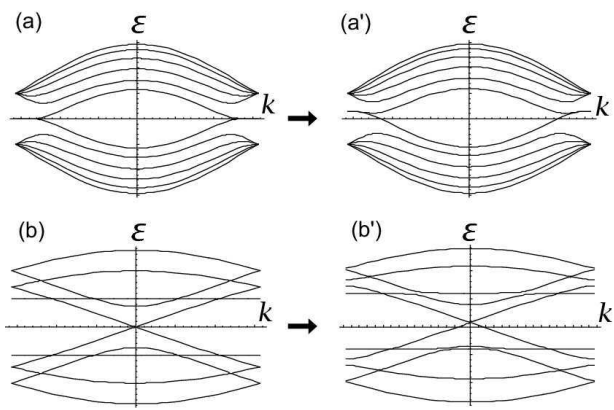}
\caption{(a) and (b); The original band structures of the $\left\langle
5,0\right\rangle $ and $\left\langle 2,1\right\rangle $ nanoribbons. (a')
and (b'); The band structures of the nanoribbon with the choice of $\protect%
\varepsilon _{\text{edge}}=\protect\varepsilon _{\text{bulk}}+0.3t$. The
inversion symmetry is broken and the Fermi energy is slightly moved from $%
\protect\varepsilon _{\text{bulk}}=0$.}
\label{EdgeBandE}
\end{figure}

Finally, the site energy is affected by the difference of electronegativity
of X. The effect is expected to be also very small because the site energy
of $\pi $ electrons is mainly determined by carbon atoms. It breaks the
inversion symmetry because the lattice of carbon atoms cannot be resolved
into two sublattices any more\cite{Lieb}. For this reason the Fermi energy
is moved from $\varepsilon _{\text{bulk}}=0$. The recalculated band
structure of zigzag and armchair nanoribbons is given by assuming a large
choice of the edge correction, $\varepsilon _{\text{edge}}=\varepsilon _{%
\text{bulk}}+0.3t$, in Fig.\ref{EdgeBandE}. The modification is small as
expected, though some of metallic armchair nanoribbons become semiconducting.

In general the transfer-energy correction must be smaller than the
site-energy correction because the former is due to a structural distortion
while the latter is due to the electronegativity. The transfer-energy
correction will be negative by considering the expansion effect near the
surface. On the other hand, the site-energy correction will be positive or
negative if the relative electronegativity of the edge atoms is positive or
negative.

\begin{figure}[t]
\includegraphics[width=0.45\textwidth]{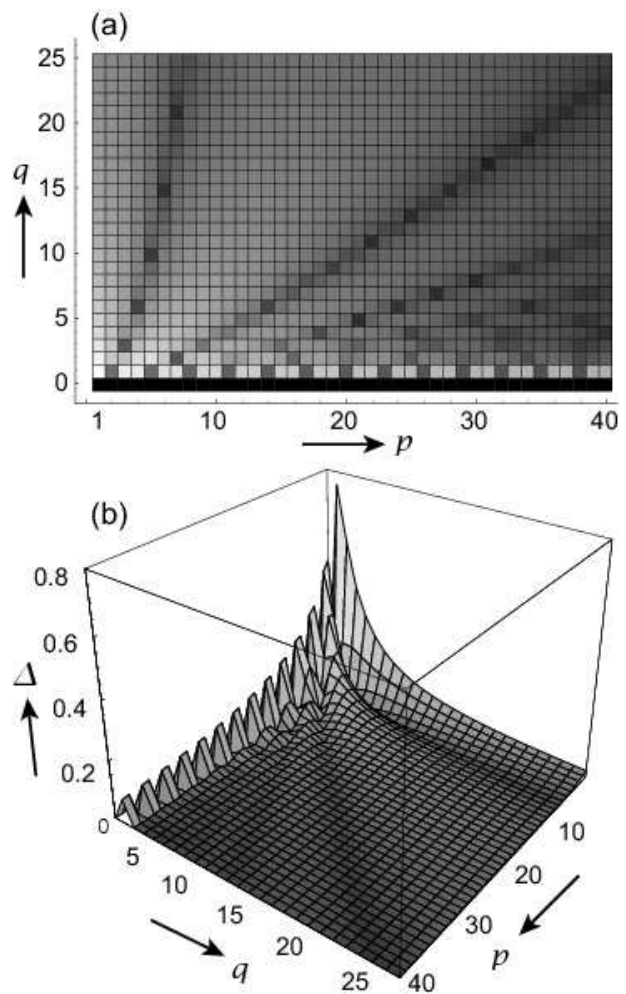}
\caption{The band gap structure of nanoribbons with the edge correction $%
\protect\varepsilon _{\text{edge}}=\protect\varepsilon _{\text{bulk}}+0.05t$
and $t_{\text{edge}}=0.99t_{\text{bulk}}$. See Fig.4 for other details.
Nanoribbons make a valley structure with stream-like sequences of almost
metallic points in the $pq$ plane. Zigzag nanoribbons remains gapless.
Armchair nanoribbons are most strongly affected by edge corrections.}
\label{Valley2}
\end{figure}

\begin{figure}[t]
\includegraphics[width=0.45\textwidth]{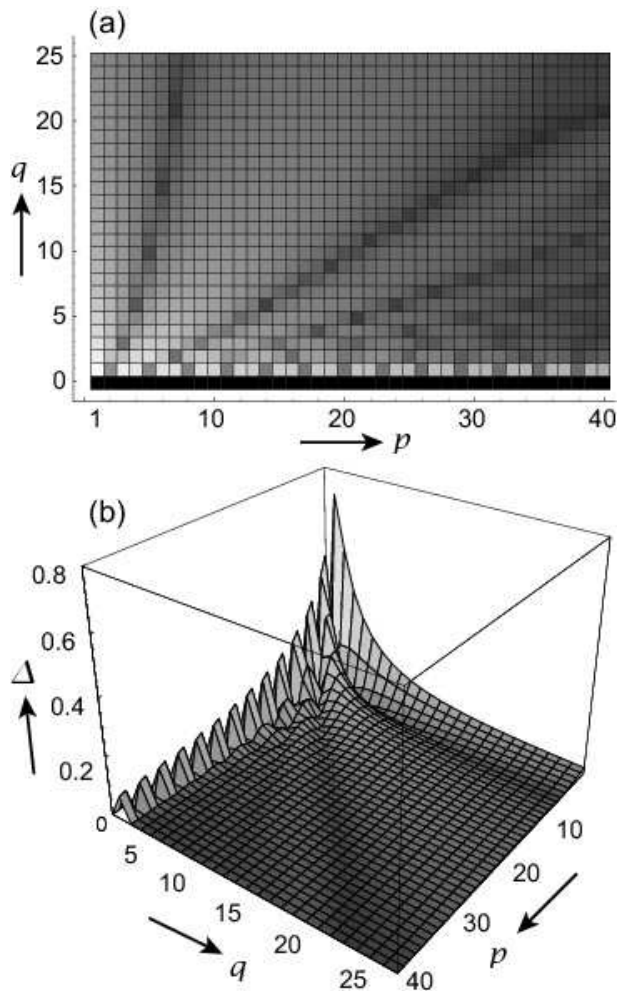}
\caption{The band gap structure of nanoribbons with the edge correction $%
\protect\varepsilon _{\text{edge}}=\protect\varepsilon _{\text{bulk}}+0.10t$
and $t_{\text{edge}}=0.95t_{\text{bulk}}$. See Fig.4 for other details.
Nanoribbons make a valley structure with stream-like sequences of almost
metallic points in the $pq$ plane. Zigzag nanoribbons remains gapless.
Armchair nanoribbons are most strongly affected by edge corrections.}
\label{Valley4}
\end{figure}

We present an overview of the band gap structure of various $\left\langle
p,q\right\rangle $ nanoribbons with the edge corrections by making a choice
of $\varepsilon _{\text{edge}}=\varepsilon _{\text{bulk}}+0.05t$ and $t_{%
\text{edge}}=0.99t_{\text{bulk}}$ in Fig.\ref{Valley2}, and by making a
choice of $\varepsilon _{\text{edge}}=\varepsilon _{\text{bulk}}+0.10t$ and $%
t_{\text{edge}}=0.95t_{\text{bulk}}$ in Fig.\ref{Valley4}. Comparing them
with Fig.\ref{Valley1}, some features of edge effects are manifest: (A) The
valley structure with stream-like sequences remains as they are. (B) All
zigzag nanoribbons remains gapless. (C) Armchair nanoribbons are most
strongly affected. (D) Edge effects are negligible for wide nanoribbons.

\section{Discussion}

\label{SecConc}

We have systematically studied and presented a gross view of the electronic
property for a wide class of carbon nanoribbons. They exhibit a rich variety
of band gaps, from metals to typical semiconductors. Zigzag and armchair
nanoribbons have electronic properties similar to nanotubes, but other
nanoribbons are quite different. It is remarkable that there exist sequences
of metallic or almost metallic nanoribbons which look like streams in valley
made of semiconductors. They approach equi-width curves for wide
nanoribbons. We have revealed a peculiar dependence of the electronic
property of nanoribbons on the width $w$. These characteristic features are
not affected strongly by edge corrections even for narrow nanoribbons.

In our analysis we have employed the nearest-neighbor tight-binding model.
It is worthwhile to calculate band gaps by more rigorous methods such as a
density-functional theory\cite{Porezag,Kusakabe,Maruyama,Higuchi}. It is
interesting to examine whether metallic points on the sequences we have
discovered remain gapless in these calculations. We also note that edge
corrections have been calculated by a tight binding density-functional
method in several cases for other materials\cite%
{Lee,Seifert,Hajnal,Seifert3,Kohler}. Needless to say it is an extremely
hard task to carry out these calculations and practically impossible to make
a systematic analysis based on them. Our results on the electronic property
of carbon nanoribbons will be useful as a guidepost for those advanced
studies.

In passing, we remark that experimental studies of nanoribbons are just in
the beginning stage\cite{Cancado,Niimi,Kobayashi} in comparison with the
study of nanotubes. This may be due to a difficulty of manufacturing and
selecting good samples, but the recent technical developing will soon solve
this problem. The band gap study of various nanoribbons presented in this
paper may be a basic step for various application of carbon nanoribbons.

The author is very thankful to Professors H. Aoki and R. Saito for various
stimulating discussions.

\end{document}